\documentclass[a4paper,11pt]{article}
\pdfoutput=1 

\usepackage{siunitx} 
\usepackage[utf8]{inputenc}
\usepackage{jinstpub}
\usepackage{pgf}
\usepackage[normalem]{ulem}
\usepackage{mathtools}
\RequirePackage{lineno}
\usepackage{color}
\usepackage{upgreek}
\usepackage{subfig}
\title{Prospects of charge signal analyses in liquid xenon TPCs with proportional scintillation in the liquid phase}

\author{Fabian~Kuger,}
\author{Julia~Dierle,}
\author{Horst~Fischer,}
\author{Marc~Schumann,}
\author{Francesco~Toschi}

\affiliation{Physikalisches Institut, Universit\"at Freiburg, 79104 Freiburg, Germany}

\emailAdd{fabian.kuger@physik.uni-freiburg.de}

\abstract{As liquid xenon TPCs increase in target mass while pursuing the direct detection of WIMP dark matter, the technical challenges arising due to their size call for new solutions and open the discussion on alternative detector concepts. 
Proportional scintillation in liquid xenon allows for a single-phase design evading problems related to the liquid-gas interface and the precise gas gap required in a dual-phase TPC.
Aside from a different scintillation mechanism, the successful detection- and analysis scheme of state-of-the-art experiments is maintained in this approach. 
We study the impact on charge signal analysis in a single-phase detector of DARWIN dimensions, where the fast timing of the proportional scintillation signal allows for the precise identification of the single electrons in the ionisation signal. 
Such a discrete electron-counting approach can lead to a better signal resolution for low energies when compared to the classical dual-phase continuous method. 
The absence of the liquid-gas interface can further benefit the S2-only energy resolution significantly. 
This can reduce the uncertainties from the scintillation and signal-detection process to a level significantly below the irreducible fluctuation in the primary ionisation. 
Exploiting the precise electron arrival time information can further allow for a powerful single vs.~multiple site interaction discrimination with 93\% rejection efficiency and 98\% signal acceptance. This outperforms the design goal of the DARWIN observatory by a reduction factor of 4.2 in non-rejected multiple site neutron events.}


\arxivnumber{2112.11844}

\begin{document}
\maketitle
\flushbottom

\section{Introduction}
\label{Sec:Intro}

The time projection chamber (TPC) is one of the most successful detector concepts for rare interaction searches in particle- and astro-particle physics.
State-of-the-art experiments aiming for the direct detection of WIMP dark matter (DM), like XENONnT~\cite{XENONnT_WIMP_2020}, LZ~\cite{LZ_WIMP_2020} and PandaX-4T~\cite{PandaX-4T:2021bab}, deploy a liquid xenon target inside a TPC instrumented with photosensors to detect light signals. 
Future DM-search experiments, like the DARWIN observatory ~\cite{DARWIN_2016}, expand their science reach to nuclear physics, like the hypothetically neutrinoless second-order weak decays of $^{136}$Xe  ~\cite{DARWIN_0vbb_2020} and $^{124}$Xe \cite{Wittweg_2020}, and the physics of solar and atmospheric neutrinos~\cite{DARWIN_2020_solarNu, newstead2020atmospheric}.
Any particle interacting with the liquid xenon target deposits energy in the form of scintillation, ionisation and heat. 
Combining the measurement of the prompt scintillation signal (S1) with the charge signal (S2), measured after the electrons are drifted to the anode, yields good energy resolution even for small energy depositions on the keV-scale~\cite{LUX_2017_LowECal}. 
Precise measurement of this charge component requires a proportional scintillation mechanism, converting each electron into a sizeable number of photons, ideally with low statistical fluctuations. 
In currently deployed dual-phase TPCs, the electrons are extracted from the liquid phase into a narrow gas gap, where electrical fields up to \SI{10}{kV/cm} are used to accelerate the electrons and stimulate scintillation of the xenon gas. 
A scintillation yield $SY$ of $(100-200)$ photons per electrons is obtained, depending on the thickness of the gas gap and the electrostatic configuration. 
A fraction $g \approx (10-15)\%$ of those photons are detected by releasing a photo electron (PE) in some of the light sensors and thus contribute to the signal.
Accordingly, the single electron gain \mbox{($G_{\rm SE}~=g\,SY$)} typically ranges from 15 to 30~PE/e$^-$~\cite{XENON100_2014_SE, XENON1T_analysis, PandaX_2016_Commissioning}. 
With the increasing diameters of liquid xenon TPCs in the current (XENONnT, LZ) and future (DARWIN) generations of direct detection dark matter experiments, maintaining geometrical precision and field uniformity in the gas gap is becoming increasingly difficult.
Sagging, the displacement of the anode wires under the electrostatic force, scales with the wire length squared and significantly reduces the gas gap thickness close to the center of the TPC. 
Thus, the effective scintillation path length and electrical field becomes $x$-$y$-dependent. 
Counteracting this sagging with increased wire tension requires stiffer structures, and is limited by the requirement of a lightweight construction to minimise background and space constraints, in particular for the anode frame.
Furthermore, the liquid-gas interface deteriorates the S1 and the S2 signal. 
Total reflection of S1 photons results in delayed detection or photon loss. 
Delayed electron extraction from the liquid surface into the gas phase reduces the observed S2, adds statistical fluctuation to the signal and can create single- or few electron signals tailing an event~\cite{XENON100_2014_SE}. 
Omitting the gas phase and stimulating the electrons to cause proportional scintillation in the liquid is a promising alternative to the established dual-phase scheme, as all
challenges related to the liquid-gas interface are avoided. 
Electrostatic sagging of the anode wires is reduced, given the symmetric electrostatic setup. 
This electroluminescence in liquid xenon has been demonstrated using thin wires \cite{LANSIART197647, Aprile_SinglePhase_2014},  exploiting the strong radial field in the vicinity of the wire surface.

In this study we explore the prospects offered by this alternative proportional scintillation mechanism to the charge signal analysis in a future liquid xenon detector, such as the proposed DARWIN experiment.
We focus on a classical TPC design with electron drift along the cylinder axis ($z$), rather than a radial setup, recently proposed~\cite{Lin_2021} and experimentally tested~\cite{SanDix_2021}. 
This retains the operational principle of past and present liquid xenon detectors and allows a direct transfer of the signal analysis scheme with the sole difference being the secondary scintillation mechanism. 
We discuss the underlying electroluminescence processes in detail in chapter~\ref{Sec:PropScint} and identify an electrode stack design for charge-to-light signal conversion as well as voltage and wire diameter configurations yielding the required $SY$.
The benefits of such a single-phase setup over state-of-the-art charge signal analyses in dual-phase detectors stem from the faster secondary scintillation process. 
The corresponding charge signal waveforms are modeled using detailed Monte Carlo techniques, presented in chapter~\ref{Sec:WFSim}. 
The distinctive signal shape per electron allows reconstruction of the arrival times of individual electrons, and therefore also an electron counting-based analysis. Neither of which are possible with similar precision in a dual-phase detector. 
The results reported in chapter~\ref{Sec:S2Analysis} focus on the impact of these two additional measures on the energy resolution, the position reconstruction and the detector capability to discriminate between single site and multiple site interactions, three key features for dark matter and rare event searches.

\section{Proportional scintillation in liquid xenon}
\label{Sec:PropScint}
In this chapter we discuss the process of photon emission after electron-xenon scattering under the influence of an electric field. 
In state-of-the-art xenon TPCs this scintillation in gaseous xenon is exploited to create a light signal proportional to the number of contributing electrons. 
We emphasise the similarities and differences of this electroluminescence in gaseous and liquid xenon and infer the number of emitted photons per electron, referred to as the scintillation yield $SY$, for a proposed application in future single-phase xenon TPCs. 

In gaseous and liquid xenon alike, the emission of a VUV photon requires an excited xenon atom Xe$^*$ to form an excimer Xe$_2^{*}$ which then dissociates~\cite{Chepel_2013}:
\begin{align}
    &\textrm{Xe} + \textrm{e}_T^- \rightarrow  \textrm{Xe}^{*} +   \textrm{e}^-_{T^\prime} \\
    &\textrm{Xe}^{*} +  \textrm{Xe} +  \textrm{Xe}  \rightarrow  \textrm{Xe}_2^{*} +  \textrm{Xe} \\
    &\textrm{Xe}_2^{*} \rightarrow  \textrm{Xe} + \textrm{Xe} + \gamma_{\rm VUV}
\end{align}

\begin{figure}[ht]
\centering
\begin{minipage}[t]{.99\textwidth}
 \centering
  \includegraphics[width=.50\linewidth]{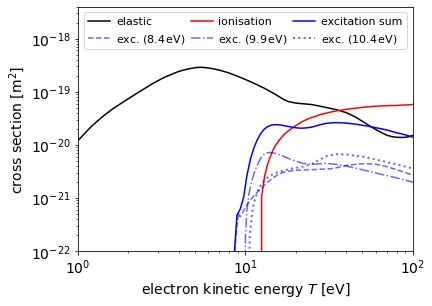}
  \captionof{figure}{Electron-xenon cross sections as a function of electron kinetic energy $T$ for elastic scattering (black), ionisation (red) and the sum of all excitations (blue), which is sub-dominant for all $T$. The three strongest contributing excitation channels are shown individually (light blue). Data from~\cite{Hayashi_DB, Hayashi_LXcat}.}
  \label{fig:CrossSections}
\end{minipage}%
\hfill
\end{figure}

Electronic xenon excitation requires an energy of \SI{8.3}{eV}. 
However, even for kinetic energies $T$ of the scattering electron above this threshold, the cross sections, shown in Figure~\ref{fig:CrossSections}, for inelastic electron-xenon scattering with xenon excitation remain sub-dominant to the elastic process~\cite{Hayashi_DB, Hayashi_LXcat}. 
While data-driven calculations of the absolute cross sections exist only for xenon in gaseous phase, the electron shell properties, thresholds and the relative strength of the processes are presumably similar in the liquid phase. 
Inelastic ionising scattering requires an energy of \SI{12.1}{eV} and becomes dominant over elastic scattering for electron energies $T > \SI{35}{eV}$. 
Due to the low excitation probability in each scatter, a large number of $\sim 10^7$ electron-xenon interactions is required to generate a sizeable signal of $\sim 10^2$ photons. In the scintillation-only regime, where electron energies don't exceed the ionisation threshold, the number of photons emitted per electron is Poisson distributed around the $SY$.  

\begin{figure}[ht]
\centering
\begin{minipage}[t]{.99\textwidth}
 \centering
  \includegraphics[width=.85\linewidth]{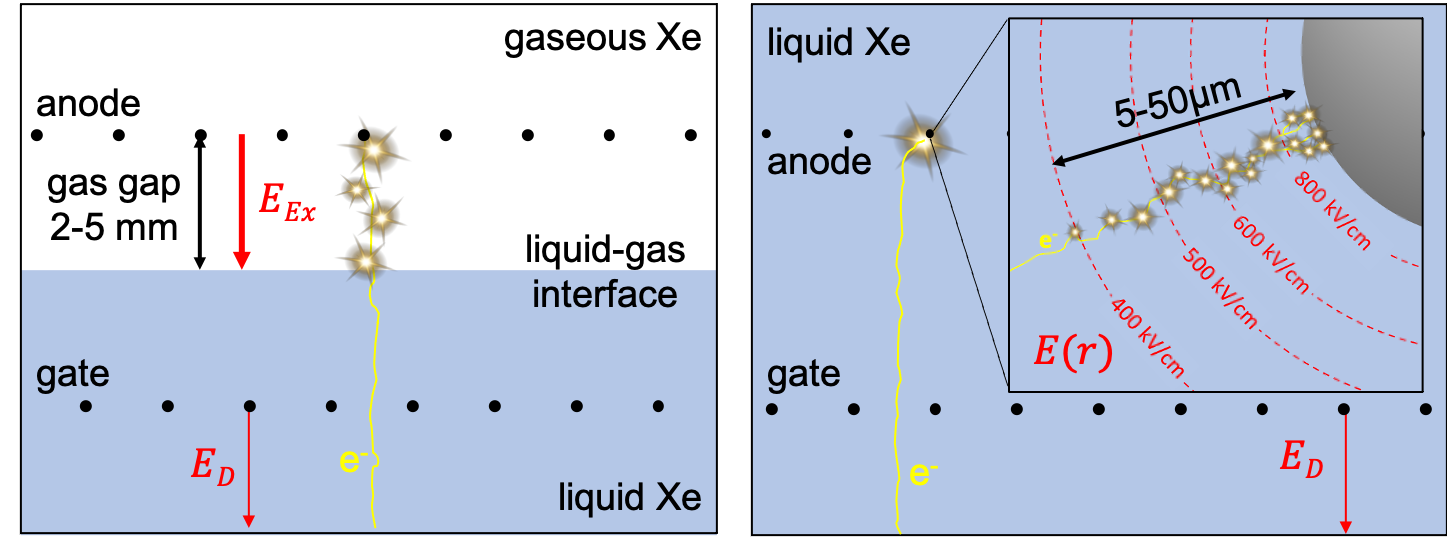}
  \captionof{figure}{Conceptual sketch of the scintillation along an electron path in the gas gap of a dual-phase detector (left) and close to a thin wire in single-phase (right).}
  \label{fig:ConceptualSketch}
\end{minipage}%
\end{figure}

The kinetic energy the electron transfers in the scattering process depends on the accelerating electrical field and the path length between collisions.
For proportional scintillation in gaseous xenon, electrical fields $E_{Ex}$ of $(8-12)\,$kV/cm are generated in a parallel-plate-like setup, allowing continuous scintillation along path lengths of $2-\SI{5}{mm}$, shown in Figure~\ref{fig:ConceptualSketch} (left).
The electron transition time through this gas gap is $\sim 1$-\SI{2}{\micro s}. 
The scintillation time distribution is in first order uniform, with deviations due to non-uniform fields along the electron path. 
In liquid xenon, the inter-atomic distance is much smaller than in gas, in the order of $2 r_{W}$, where  $r_{W} \sim \SI{0.2}{nm}$ is the Van-der-Waals radius of a xenon atom. 
The electron drift path between scatters is therefore smaller and the local field strengths must exceed several \SI{100}{kV/cm} to allow for xenon excitation with non-vanishing probabilities. 
On the other hand, much shorter electron paths suffice to ensure a large number of scatters. 
Such a field configuration can be realised in the vicinity of thin wires, shown in Figure~\ref{fig:ConceptualSketch} (right), where the radial $E(r) \propto 1 / r$ behaviour allows for strong fields at moderate bias voltages, provided the wire radius is sufficiently small. 
Accordingly, the scintillation path length is reduced to the order of the wire radius and the corresponding electron transition time is a few ns. 

For stronger fields the electrons' energy is more likely to be above the ionisation threshold, rendering this process dominant over xenon excitation.
While this reduces the excitation probability per scatter, the additional electrons freed in ionisation contribute to the total scintillation yield. 
Additional excimer formation by recombination~\cite{Chepel_2013} is suppressed by the strong electric field, and thus contributes little. 
While the net growth in $SY$ is beneficial for signal detection, the statistical fluctuations in electron avalanche formation will lead to a larger width $\sigma_{SY}$ in the $SY$ statistics. 
The  single electron response spectrum, which is Poisson distributed in the scintillation-only regime, will be distorted by the broader Polya-shaped spectrum characteristic of electron avalanche formation~\cite{Alkhazov:1970fx}. 
The spectral broadening is limited for small charge multiplication factors, where the single or few ionisation processes occur only very close to the wire surface and the liberated electron contributes little, if at all, to the scintillation. 

The degradation of energy resolution in transition from the scintillation-only to the charge multiplication regime has been observed using electrons from an $\alpha$-source to create proportional scintillation in liquid xenon using \SI{5}{\micro m} and \SI{10}{\micro m} diameter wires \cite{Aprile_SinglePhase_2014}. 
The resolution loss is limited for charge multiplication factors $f_Q < 3$ and becomes significant above. 
The authors of \cite{Aprile_SinglePhase_2014} describe their measurements using a data-driven exponential and linear model for charge multiplication $\Delta N_e$ and scintillation $\Delta N_\gamma$, respectively, per incremental spatial step $\Delta \vec{x}$:
\begin{align}
    \Delta N_e &= N_e \theta_0 \exp{\left(-\frac{\theta_1}{E(\vec{x}) - \theta_2}\right)} \vert \Delta \vec{x} \vert     \label{eq:FqModel} \\
    \Delta N_\gamma &= N_e \theta_3 \left(E(\vec{x}) - \theta_4\right) \vert \Delta \vec{x} \vert\:,
    \label{eq:SYModel}
\end{align}
where $E(\vec{x})$ is the electrical field strength at spatial position $\vec{x}$ and $\theta_i$ are model parameters. Note that equations \eqref{eq:FqModel} and ~\eqref{eq:SYModel} hold only for $E(\vec{x}) > \theta_2$ and $E(\vec{x}) > \theta_4$, respectively.
They find field thresholds of $\theta_4 \approx \SI{400}{kV/cm}$ for scintillation and $\theta_2 \approx \SI{700}{kV/cm}$ for ionisation. 
The quoted uncertainty on the model parameters $\theta_i$ is up to 30\% and the scintillation turn-on at comparatively low fields is not well described by the model, as noted by the authors. 
A revised model with accurately determined parameters is mandatory for the precise quantification of charge-to-light conversion in liquid xenon. This calls for systematic experimental studies and a more detailed model based on the microscopic processes. 
However, the empirical model in~\cite{Aprile_SinglePhase_2014} provides a first approximation for the achievable $SY$ and charge multiplication in a given electrostatic setting and is used throughout this study. 
Since this study extends to thicker wire diameters, the self-shadowing by photon absorption on the wire surface has been investigated assuming isotropic light emission and no surface reflectivity. 
The presumed increase in photon loss with increasing wire diameter is found to be limited to 2\%, comparing \SI{10}{\micro m} with \SI{50}{\micro m} wire diameter.  
The difference in photon detection efficiency between top and bottom sensors, observed and attributed to self-shadowing in~\cite{Aprile_SinglePhase_2014}, is mitigated in this study by the large size of the TPC.

\begin{figure}[ht]
\begin{minipage}[t]{.48\textwidth}
 \centering
  \includegraphics[width=.99\linewidth]{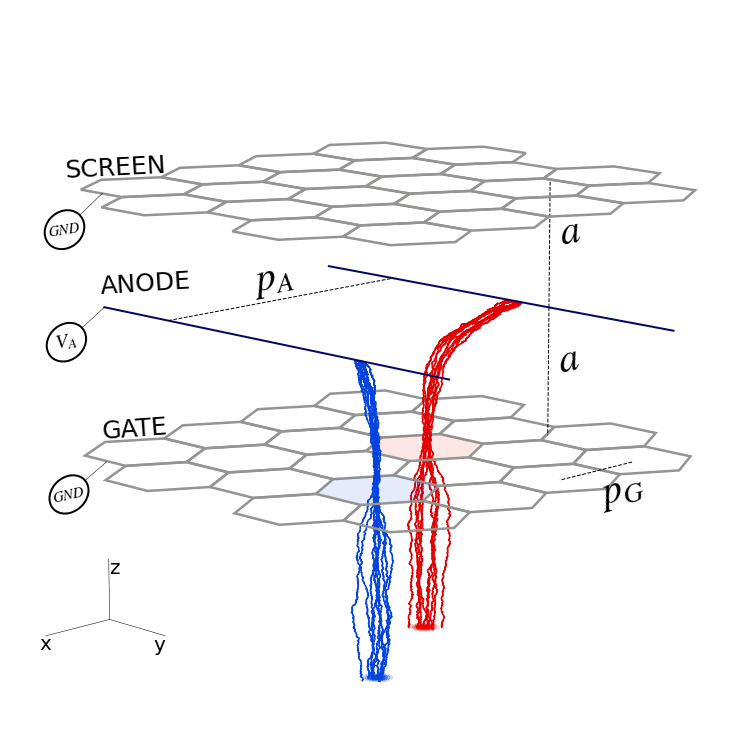}
  \captionof{figure}{Electrode stack setup with the anode wire grid and etched electrodes with hexagonal patterns: the gate (bottom) and the screen (top), all immersed in liquid xenon. Drift lines correspond to two electron clouds, each funnelled through a different gate hexagon, leading to an $x$-$y$-dependent drift time with distinct populations (Figure~\ref{fig:HexGrid_timing}, bottom). }
  \label{fig:Setup_figure}
\end{minipage}%
 \hfill
\begin{minipage}[t]{.48\textwidth}
 \centering
  \includegraphics[width=.99\linewidth]{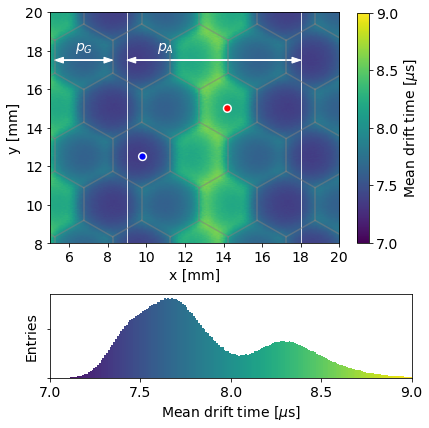}
  \captionof{figure}{Electron drift time through the electrode stack as a function of the $x$-$y$ starting position \SI{10}{mm} below the gate. Hexagonal grid structure (grey) and anode wire (white) are indicated. The blue and red dots indicate the transit position of the drifting electrons in Figure~\ref{fig:Setup_figure}. The distribution (bottom) features two populations, of faster and slower electrons.}
  \label{fig:HexGrid_timing}
\end{minipage}%
\end{figure}

\begin{figure}
\centering
\begin{minipage}{.48\textwidth}
 \centering
  \includegraphics[width=.99\linewidth]{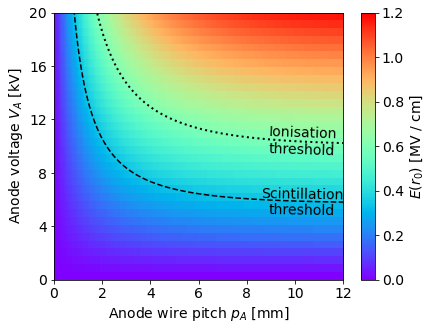}
  \captionof{figure}{Electrical field strength $E$ at the surface of a \SI{50}{\micro m} diameter wire grid anode at $a = \SI{5}{mm}$ distance to gate and screen as function of the anode pitch $p_A$ and the applied voltage $V$. Field thresholds~\cite{Aprile_SinglePhase_2014} for scintillation (dashed) and ionisation (dotted) are shown for reference. }
  \label{fig:Field_pitchratio}
\end{minipage}%
\hfill
\begin{minipage}{.48\textwidth}
 \centering
  \includegraphics[width=.99\linewidth]{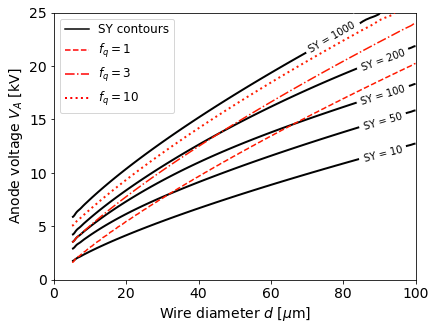}
  \captionof{figure}{Contours of the scintillation yield $SY$ for a range of wire diameters and anode voltages, overlayed with the charge multiplication factor $f_Q$ contours (red). Each contour crossing identifies the minimal voltage and required wire diameter to reach the corresponding $SY$ with limited $f_Q$.}
  \label{fig:SY_d-V-plane}
\end{minipage}%
\end{figure}

Following the successful detection principle of past and existing dual-phase TPC experiments, proportional scintillation in liquid should be realised throughout the $x$-$y$-plane of a future single-phase detector, with $z$ being the drift direction. 
We propose a wire grid anode, depicted in Figure~\ref{fig:Setup_figure}, with thin wires on high positive potential with equal distance $a$ from two grounded electrodes.
The gate electrode on the bottom electrostatically separates the high-field region from the TPC drift region. The top screen symmetrises the field, maximising the field strength at the anode for a given anode voltage $V_A$. 
Furthermore a symmetric field limits the net electrostatic force on the anode and the resulting wire displacement in $z$. 
A smaller inter-electrode distance $a$ provides stronger fields close to the anode and thus higher $SY$ at a given $V_A$. It likewise increases the surface field on the gate and screen, which is observed to cause spontaneous electron emission~\cite{Pixey_2021}. 
The gate and screen surface fields can be reduced by smoothing their surfaces~\cite{Tomas:2018pny} or decreasing the pitch $p_G$, at the cost of reduced optical transparency. 
It is generally lower on mesh-like structures, such as the etched hexagonal meshes used in XENON1T~\cite{XENON1T_2017}, compared to wire grids with equal fill factor. Hexagonal meshes with $p_G$ = \SI{3}{mm} are used throughout this study. 
Increasing the anode wire pitch $p_A$ at constant $V_A$ and $a$ yields stronger fields around the anode wires, as shown in Figure~\ref{fig:Field_pitchratio}, and higher $SY$. 
However, it degrades the $x$-$y$-position resolution of the detector by causing increased electron displacement before scintillation. 
The corresponding additional drift time, shown in Figure~\ref{fig:HexGrid_timing}, contributes to the electron arrival time pattern and also affects $z$-position reconstruction and discrimination of single site vs.~multiple site interactions, as discussed in chapter~\ref{Sec:z_Reco} and \ref{SS-MS_Disc}. 
To minimise these $x$-$y$-dependencies the anode wire pitch is selected as a multiple of the gate pitch $p_G$. While an increased $p_G$ combined with a 90$^\circ$-rotated alignment of the electrodes decreases the drift time spread, it requires an increased voltage to obtain the same local field strength around the anode wire. 
Accounting for the experimental challenges arising at these voltages, we conservatively present the case with a lower voltage requirement. 
We find that a setup with $p_A =  1.8 a = 3 p_G = \SI{9}{mm}$, the configuration in Figures~\ref{fig:Setup_figure} and \ref{fig:HexGrid_timing}, balances well between these aspects. 
The electron trajectories, shown in Figure~\ref{fig:Setup_figure}, are well defined throughout the $x$-$y$-plane, provided gate and anode are aligned with sufficient precision. 

Based on the above discussed model~\cite{Aprile_SinglePhase_2014} we calculate scintillation yield and average charge multiplication factor $f_Q$ for $a= \SI{5}{mm}$ and a range of anode wire diameters $d$ and applied voltages, as shown in Figure~\ref{fig:SY_d-V-plane}. 
The crossing point of the contours indicate the minimal voltage and wire diameter required to obtain the corresponding $SY$ with limited charge multiplication to $f_Q$.
Avoiding charge multiplication completely limits the obtainable $SY$ to 50 photons per electron for $d=\SI{35}{\micro m}$ wires ($V_A = \SI{8.4}{kV}$) or 100$\,\gamma\,/\,e^-$ for \SI{70}{\micro m} wires ($V_A = \SI{14.8}{kV}$). 
Constraining to an average of two ionisation processes ($f_Q = 3$), where resolution degradation remains limited, as suggested by the data in~\cite{Aprile_SinglePhase_2014}, $SY=100\,\gamma\,/\,e^-$ could be realised with anode wires as thin as \SI{10}{\micro m} ($V_A = \SI{5.0}{kV}$). 
The lowest voltage configuration to achieve $SY=200\,\gamma\,/\,e^-$ with $f_Q \leq 3$ is \SI{14.3}{kV} on \SI{50}{\micro m} diameter wires.
$SY > 1000\,\gamma\,/\,e^-$ requires the formation of an electron avalanche and $f_Q > 10$.  
A reduced inter-electrode distance $a$, introduced by sagging of the gate and screen, increases $E(r_0)$ and accordingly the $SY$ and $f_Q$. For a \SI{50}{\micro m} wire operated with $V_A = \SI{12.5}{kV}$ ($SY = 100$), a distance reduction by 10\% (20\%) increases the $SY$ by 13\% (28\%). 
This $SY$ increase shows a mild dependence of $\pm 1\%$ ($\pm 2\%$) on the wire diameter. 
Increasing $a$ limits the relative effect of a given absolute electrode deformation, but requires higher voltage to obtain the same surface fields on all electrodes.

\section{Waveform simulation for a single-phase TPC}
\label{Sec:WFSim}
A detailed simulation of the time sampled light sensor response to the electron induced scintillation signal, referred to as  S2 waveform, is key to explore the prospects of charge signal analysis methods for single-phase TPCs. 
The proportional scintillation region is modeled with the anode grid design concluded on in chapter~\ref{Sec:PropScint}, with the \SI{50}{\micro m} wire configuration allowing for $SY = 200\,\gamma\,/\,e^-$ with limited charge multiplication. 
Aiming for an application in a next generation dark matter experiment, the overall detector is modeled according to the design of the proposed DARWIN observatory~\cite{DARWIN_2016}, with TPC dimensions of \SI{2.6}{m} in diameter and height and the light sensor arrangement taken from~\cite{DARWIN_0vbb_2020}. 
The assumed time sampling of \SI{100}{MHz} (\SI{10}{ns} / bin) matches the readout electronics used in XENONnT and LZ. 
We note that an increased sampling rate would further benefit the electron counting and arrival time reconstruction discussed in chapter~\ref{Sec:S2Analysis}, up to the limit where sampling uncertainty becomes negligible compared to PMT intrinsic time resolution caused by transition time spread. 

The electron cloud formed by a single energy deposition is assumed to be point-like compared to the anode pitch. While for larger energies, such as in $0\nu\beta\beta$ searches, the electron thermalisation path can extend over several millimeters~\cite{DARWIN_0vbb_2020}, the electron-ion pairs from low-energy interactions are confined into much smaller volumes, justifying this assumption. 
These electron clouds diffuse along their drift path towards the gate as discussed in detail in~\cite{LI2016160}. Within the bulk of the liquid xenon target we assume a uniform drift field of $E_D = \SI{250}{V/cm}$ and compare this baseline choice to higher (\SI{500}{V/cm}) and lower (\SI{100}{V/cm}) field strength. For the drift along a path length $z$ we apply the corresponding analytic diffusion approximation for each direction of movement:
\begin{equation}
    \sigma_{L / T} = \sqrt{\frac{2 D_{L / T} \, z}{v_D}} 
    \quad \textrm{  (longitudinal / transverse)}
    \label{eq:Diffusion}
\end{equation}
with drift velocity $v_D = \SI{1.6}{mm \per \micro s}$ and diffusion constants $D_L = \SI{2.4e-3}{mm\squared \per \micro s}$~\cite{Hogenbirk_2018} and $D_T = \SI{5.5e-3}{mm\squared \per \micro s}$~\cite{PhysRevC.95.025502}  for $E_D = \SI{250}{V \per cm}$. The longitudinal diffusion randomises the arrival time of individual electrons at the gate according to a normal distribution with $\sigma_{t} = \sigma_{L} / v_D$.
The transverse diffusion causes electron spread in the $x$-$y$-plane according to $\sigma_{T}$ for each direction. 
For $z \gtrapprox \SI{2}{m}$ the transverse diffusion $\sigma_{T}$ reaches the scale of the hexagonal mesh cells of \SI{3}{mm}, thus electron clouds are split and transit the gate electrode at different positions. 
The spatial extent remains smaller than the anode wire pitch of \SI{9}{mm}, thus electrons are typically collected  on one, occasionally on two wires.  
For each simulated event the $x$-position modulo $p_A$ is randomly selected. The electron arrival positions at the gate then follow the normal distribution with $\sigma_{T}(z)$.
The transition region to the non-uniform field below the top electrodes is modeled in detail using COMSOL~\cite{comsol}, starting $\SI{10}{mm}$ below the gate where the field starts to divert from uniformity. The $x$-$y$-dependent contribution of this last step to the electron arrival time at the anode wire is shown in Figure~\ref{fig:HexGrid_timing}. 
The distribution features two populations of electrons collected faster or slower at the anode, depending on the gate mesh cell they transit through~\cite{PyCOMes}. 

For each electron approaching an anode wire we model the number of emitted photons with a Poisson distribution centred at the envisaged scintillation yield of $SY = 200\,\gamma\,/\,e^-$. Their assumed detection probability $g = 0.15$ is accounted for with a binomial model. 
The transition time of the electron through the region of sufficiently high field for scintillation has been simulated and found to be approximately \SI{2}{ns} and \SI{10}{ns} for wires with \SI{10}{\micro m} and \SI{50}{\micro m} diameter, respectively. 
The majority of the excitations occur within the last \SI{0.5}{ns} and \SI{2}{ns}, respectively, before the electron arrives at the wire surface. Accordingly, the detection time of each photon is dominated by the xenon de-excitation time and the photon propagation time to the light sensor. 
The xenon de-excitation of the singlet $^1\Sigma_u^+$ (BR = 5\%) and the triplet $^3\Sigma_u^+$ (95\%) state occurs with decay constants of \SI{2.2}{ns} and \SI{27}{ns}, respectively~\cite{Chepel_2013}. 
The slower contribution from photon emission after recombination with $\tau = \SI{34}{ns}$ is suppressed by the low number of ionisation scatters and the strong electrical fields. 
The time distribution of photon propagation through the detector has been modeled in a detailed Chroma~\cite{Seibert2011FastOM, Luke_Igor_PC} simulation with randomised $x$-$y$-position and photon emission angle.
Approximately 40\% of the photons are promptly detected in one of the top array PMTs. 
Their distribution follows an exponential with detection time constant $\tau_\textrm{prompt} \approx \SI{0.2}{ns}$. 
The largest fraction undergoes multiple Rayleigh scattering with mean scattering length $l_\textrm{Ray} = \SI{35}{cm}$~\cite{grace2017index} before being either detected or absorbed. This delayed contribution is modeled by summing over three exponential components, with the result that only 20\% of the total photon count follows a detection time constant larger than \SI{10}{ns}.
Variation of the reflectivity of the PTFE panels $R_\textrm{PTFE} \in [0.95, 0.99]$ surrounding the TPC and the LXe photon absorption length $L_\textrm{abs} \in [20, 50]$\,m in the simulation affects the overall light collection efficiency, but has minor impact on the photon arrival time distribution. This is due to the large TPC diameter and height compared to $l_\textrm{Ray}$.
The light sensor response is modeled assuming the Hamamatsu PMT R11410 used in the XENONnT and LZ experiments. We account for a $p_\textrm{DPE}$ = 20\% probability of a double photoelectron emission from the PMTs photocathode as measured in~\cite{LOPEZPAREDES201856} as well as for a transition time spread delaying the PMT response~\cite{Antochi_2021}. The occurrence of dark counts of a rate of \SI{24}{Hz}~\cite{XENON1T_2017} per channel is added on top of a white noise baseline.

\begin{figure}[h]
\centering
\begin{minipage}[t]{.48\textwidth}
 \centering
  \includegraphics[width=.99\linewidth]{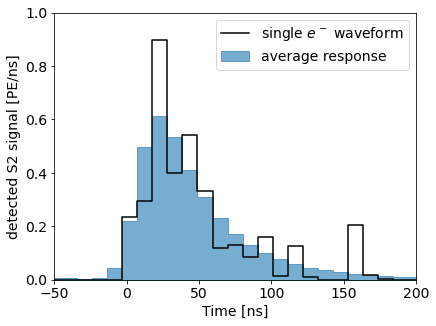}
  \captionof{figure}{Exemplary (black) and average (blue) simulated waveform for a single electron signal assuming $(SY \, g) = 30$ detected $\gamma\,/\,e^-$ and \SI{100}{MHz} sampling rate. The template signal decays exponentially with a time constant of \SI{37}{ns}.}
  \label{fig:SE_Template}
\end{minipage}
\hfill
\begin{minipage}[t]{.48\textwidth}
  \centering
  \includegraphics[width=.99\linewidth]{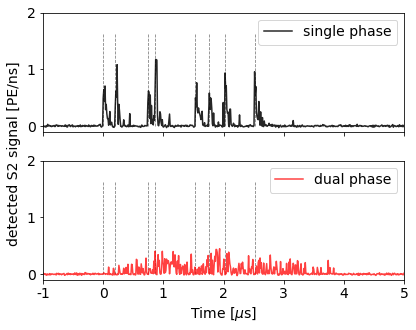}
  \captionof{figure}{Simulated waveforms of the same eight-electron event in single-phase (top) and dual-phase configuration (bottom). The true electron arrival times (dashed grey) coincide with the peaks  in the single-phase waveform. }
  \label{fig:S2_example}
\end{minipage}
\end{figure}

The simulated, detected light signal per electron is shown in Figure~\ref{fig:SE_Template}. 
The response timing is characterised by the exponential decay of the signal with a \SI{37}{ns} time constant. 
This timing is predominantly caused by the triplet state decay time with additional broadening contributions from delayed photon detection, the PMT transition time spread and the limited time resolution of the \SI{100}{MHz} sampling. 
The prolonged tail in the average response template can be attributed to photons detected after a long random walk.
The single-phase waveforms from few electron signals - an example is shown in Figure~\ref{fig:S2_example} (top) - feature characteristic sharp signal peaks for each  electron. In the more continuous S2 waveforms characteristic of a dual-phase setup, shown for the same event in Figure~\ref{fig:S2_example} (bottom), this distinct feature is lost due to the long scintillation time per electron.

\section{Charge signal analyses}
\label{Sec:S2Analysis}

\begin{figure}[htb]
\begin{minipage}[c]{0.5\textwidth}          
    \includegraphics[width=\textwidth]{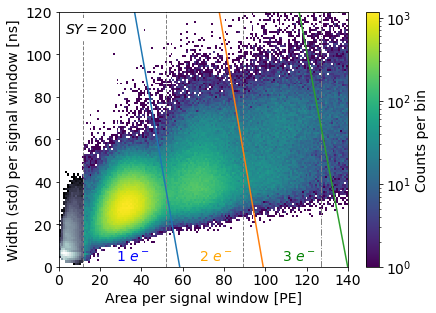}
\end{minipage}
\begin{minipage}[c]{0.5\textwidth}  
    \includegraphics[width=\textwidth]{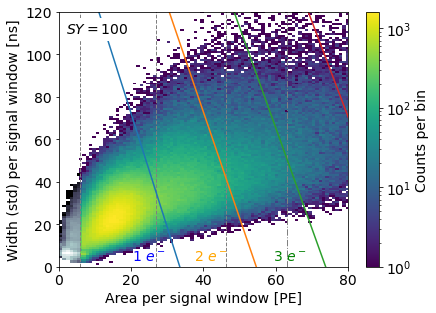}
\end{minipage}
    \captionof{figure}{The temporal signal width vs.~integrated area per signal window sampled from 20 electron events with $SY =  200\,\gamma\,/\,e^-$ (left) and $100\,\gamma\,/\,e^-$ (right). Coloured lines indicate discrimination thresholds between single ($1~e^-$) and multiple ($2~e^-, 3~e^-,$...) electron signals. Cuts in area (dashed) are shown for reference and only applied to reject small signals such as coincident dark counts (grey).} 
    \label{fig:ElectronCounting_Area-Width}
\end{figure}

The characteristic peak shape of an electron signal allows for a novel approach in charge signal analysis exploiting accurate electron counting and precise reconstruction of the electron arrival times. 
Within this study different waveform analysis algorithms have been tested using simulated single-phase charge signals. Best performance for both electron counting and time reconstruction has been obtained with a three step-approach: defining signal windows within the waveform, splitting those featuring a multiple peak structure, and finally counting the electrons within each signal window. 
The search for signal windows within a waveform uses a time-over-threshold condition for consecutive time bins. 
The threshold and the number of required consecutive bins is tuned such that single-photon PMT responses will prolong a signal window, but not trigger a new signal. 
This provides an efficient rejection of single channel dark count signals. 
Signal windows longer than \SI{150}{ns} are tested for a multiple-peak structure and split if several peaks are detected. 
The number of electron signals in each signal window is finally determined using the signal time width vs.~signal area space, shown in Figure~\ref{fig:ElectronCounting_Area-Width}.
For the event sample with $SY=200\,\gamma\,/\,e^-$ in Figure~\ref{fig:ElectronCounting_Area-Width} (left) the miscounting rate decreases by about 30\% using the area-width cut (coloured lines) compared to a threshold in signal area only (dashed lines). 
Using non-linear discrimination lines showed a comparably small additional gain in counting accuracy, but increases the calibration complexity significantly.

The counting accuracy is limited by the overlap of the single, double, triple etc. electron populations in the area-width space. 
The overlap in area is driven by the statistical fluctuation in the scintillation and photon detection process. 
Accordingly, a lower scintillation yield, resulting in a higher relative fluctuation, causes less well separable populations. 
This can be seen in direct comparison of the event samples in Figure~\ref{fig:ElectronCounting_Area-Width} generated with $SY = 200\,\gamma\,/\,e^-$ (left) and $100\,\gamma\,/\,e^-$ (right). 
Further increasing the scintillation yield reduces this overlap in area and yields a more precise electron counting. 
An overlap in width is unavoidable considering multiple electron signals in perfect time coincidence. 
The efficiency of splitting signals from two electrons arriving at a similar time depends on the algorithm applied and is limited by the time scale of a single electron signal, shown in Figure~\ref{fig:SE_Template}. 
The fluctuations of this waveform around the average template are smaller for a larger scintillation yield, which allows for better discrimination by signal width. 
Increased electron cloud diffusion increases the average time between two electron signal and reduces the coincidence probability.
Accordingly, the number of electrons in events occurring at larger $z$ can be counted with slightly higher accuracy. 
For all results presented here the $z$-position is uniformly sampled throughout the TPC, unless explicitly noted.

The electron arrival times are reconstructed using a matched filter to compare the waveform with the single electron response template. 
Electron arrival times at the anode wire are measured with a $1\sigma$-resolution of \SI{10}{ns}. 
A slight increase to $\sigma \leq \SI{15}{ns}$ is observed for an increasing number of electrons as well as for shorter drift lengths. Both increase the fraction of multiple electrons per signal window and result in an increased miscounting rate and herefore less precise reconstruction of the electron arrival times. 
The resolution limit from the \SI{100}{MHz} time sampling and corresponding \SI{10}{ns} time binning contributes to the overall time resolution, but is sub-dominant. 
Accordingly, the electron time measurement could benefit more from a refined time reconstruction algorithm than an increased sampling rate. 

As an alternative to the analytic approach described above, we applied convolutional neural networks optimised to analyse time serial data using the sktime-dl software package~\cite{sktime_2019}. 
The results obtained are similar for electron counting and slightly worse for arrival time reconstruction. 
Advanced architecture and training strategies could further improve the method. 

In the following we focus on the analytically obtained results comparing $G_{\rm SE} = $15~PE/e$^-$ and 30~PE/e$^-$ and assuming anode wires with \SI{50}{\micro m} diameter and uniform $SY$ throughout the $x$-$y$-plane.
The geometric and electrostatic configuration to reach the required $SY$, discussed in chapter~\ref{Sec:PropScint}, has negligible impact on the results presented below. 
The choice of the wire diameter has small impact on the photon losses due to the wire shadow and the photon timing, as discussed in the chapters~\ref{Sec:PropScint} and ~\ref{Sec:WFSim}, respectively.

\subsection{Electron counting and S2-only energy resolution}
\label{Sec:E_Counting}

\begin{figure}[htb]
\centering
\begin{minipage}[t]{.48\textwidth}
  \centering
  \includegraphics[width=.99\linewidth]{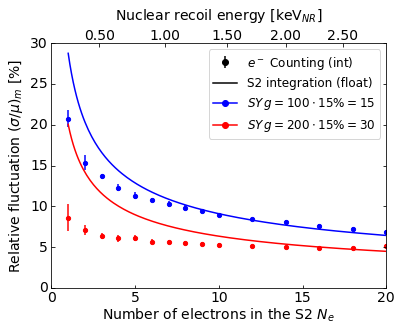}
  \captionof{figure}{Relative fluctuation in the charge signal measurement $(\sigma/\mu)_{\rm m}$ as a function of the number of electrons causing the signal with $SY = 100$ (blue) or $200\,\gamma\,/\,e^-$ (red). Electron counting (dots) is compared to the non-integer measure by integrated S2 area (lines). Conversion to recoil energy uses NEST~\cite{NEST} and assumes $E_D = \SI{250}{V/cm}$ and 100\% electron survival.}
  \label{fig:ElectronCounting_Resolution}
\end{minipage}%
\hfill
\begin{minipage}[t]{.48\textwidth}
  \centering
  \includegraphics[width=.99\linewidth]{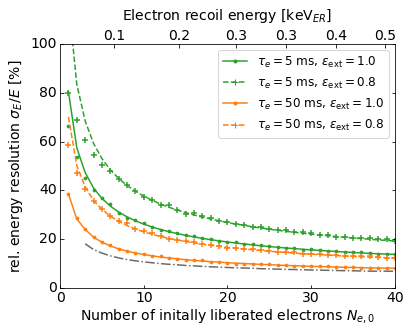}
  \captionof{figure}{Relative energy resolution $\sigma_E / E$ in charge only analysis estimated with equation \eqref{eq:Resolution_Combi} (lines) compared to full simulation with exact statistical treatment. Data is shown for $\epsilon_{\rm ext} = 0.8$ (crosses, dashed) and 1.0 (dots, solid) and $\tau_e$ of \SI{5}{ms} (green) and \SI{50}{ms} (orange). The fluctuation in ER electron liberation (grey) and corresponding energy scale is calculated by NEST~\cite{NEST}.}
  \label{fig:E_res_Comp}
\end{minipage}
\end{figure}

In charge signal (S2) analyses the signal size is typically measured in non-integer multiples of photo electrons [PE] by integrating the signal waveform $W\! F(t)$ which is converted to [PE~/~ns]:
\begin{equation}
    {\rm S2}[{\rm PE}] = \int_{t_{\rm start}}^{t_{\rm end}} W\! F(t) \,dt 
    \qquad {\rm and} \qquad {\rm S2}[{\rm e}^-]_{\rm float} = \frac{S2 [\rm PE]}{G_{\rm SE}}\:.
\end{equation}
The equivalent as continuous non-integer approximation of the true number of electrons contributing to the signal $N_e$ is approximated by division by the single electron gain $G_{\rm SE}$. 
Electron counting in single-phase provides a discrete integer measure of the number of charge signal quanta $S2 [{\rm e}^-]_{\rm int}$. 
In either case $(\sigma/\mu)_{\rm m}(N_e)$ quantifies the relative fluctuation in the S2 generation by proportional scintillation of $N_e$ electrons and the measurement of this signal S2[e$^-$], either as float or integer. In the continuous S2 measurement the fluctuations arise from the statistical nature of the proportional scintillation and the photon detection processes, described by single electron gain $G_{\rm SE}$ and its spread $\sigma_{G_{\rm SE}}$. It can be parameterised by 
\begin{equation}
    \left(\frac{\sigma}{\mu}\right)_{\rm m} (N_e) = \frac{\sigma_{G_{\rm SE}}}{G_{\rm SE}} \frac{1}{\sqrt{N_e}}\:.
\end{equation}
In the integer measure in a single-phase detector these statistical fluctuations in the scintillation and detection process don't contribute directly to the precision of the measurement, but instead limit the performance of the electron counting algorithm, as discussed above. Figure~\ref{fig:ElectronCounting_Resolution} compares $(\sigma/\mu)_{\rm m}$ between both approaches for a $SY = 100$ and 200 photons per electrons.
For $SY = 200\,\gamma\,/\,e^-$, the signal window based counting yields a measurement fluctuation of \mbox{$(\sigma/\mu)_{\rm m} \approx 5-\SI{6}{\%}$}, an improvement compared to the S2 integration for $N_e \leq 15$.  
For a reduced scintillation yield of $SY = 100\,\gamma\,/\,e^-$, the discrimination power between zero and single (and between single and double, etc.) electron signals decreases.
The relative fluctuation in the S2 measurement, shown in blue in Figure~\ref{fig:ElectronCounting_Resolution}, is therefore larger and only better than area integration for $N_e \leq 10$. 
In any case the benefit of the electron counting approach is limited to the low $N_e$ range, with the upper bound depending on the performance of the counting algorithm, the scintillation yield and the diffusion of the electron cloud.

To estimate the energy resolution $\sigma_E / E$ in the charge signal analysis, comparing single- to dual-phase, we consider the statistical fluctuations from three processes: First, the electron liberation in the primary interaction, second, electron losses during drift and in dual-phase extraction into the vapour, and third, the proportional scintillation and S2 measurement. 
The number of free electrons leaving the site of the primary interaction $N_{e,0}$ is determined by the deposited energy $E$ and the type of the interaction depositing this energy, i.e.~a nuclear recoil or an electron recoil. 
It depends on the local electric field $E_D$, which affects the recombination probabilities~\cite{Chepel_2013}. We use the Noble Element Simulation Technique (NEST) software package~\cite{NEST} to calculate the mean number of electrons liberated during the interaction $\bar{N}_{e,0}$ and the associated relative fluctuations  $(\sigma/\mu)_{\rm l}$.
Not all of those $N_{e,0}$ electrons which are initially liberated contribute to the S2 signal of $N_e$ electrons. The average ratio  $\bar{N}_e / \bar{N}_{e,0}$ equals the mean electron survival probability $P_s$. The main sources of electron loss or delay are attachment to electro-negative impurities along the drift path and, in the case of a dual-phase detector, non-prompt extraction from the liquid into the gas phase~\cite{Gushchin_1979, Xu:2019dqb}. 
The probability $P_D$ for one single electron to survive the drift process with velocity $v_D$ along a path length $z$ depends on the electron lifetime $\tau_e$, a measure of the xenon purity:
\begin{equation}
    P_{D} = \exp \left(-\frac{z}{v_D \tau_e}\right)
\end{equation}
In a dual-phase TPC, where electrons must cross the liquid-gas interface before scintillation, this drift survival probability $P_{D}$ must be multiplied with the extraction efficiency $\epsilon_{\rm ext}$. 
In absence of the gas phase in a single-phase TPC, this potential loss process is avoided completely. 
Considering the drift and extraction of $N_{e,0}$ electrons as binomial process in the Poisson approximation, the fluctuation in electron survival can be estimated as: 
\begin{equation}
   \left(\frac{\sigma}{\mu}\right)_{\rm s} (N_{e,0}) = 
   \frac{\sqrt{N_{e,0} (1 - P_s)}}{N_{e,0} P_s} = 
   \frac{1}{\sqrt{N_{e,0}}} \frac{\sqrt{1 - P_{D} \epsilon_{\rm ext}}}{P_{D} \epsilon_{\rm ext}}. 
\end{equation}
Note that the relative fluctuation in electron survival is determined by the absolute fluctuation in electron loss $\sqrt{N_{e,0}\,(1 - P_s)}$ normalised to the mean number of surviving electrons $N_{e,0}\, P_s$. 
Accordingly, $(\sigma/\mu)_{\rm s}$ converges to 0 for small loss probability or large $N_e$.

Based on the fluctuations in electron liberation ($l$), survival ($s$) and measurement ($m$), the relative energy resolution in a charge signal only measurement can be approximated by:
\begin{equation}
    \left(\frac{\sigma_E}{E}\right)_{S2}  = \sqrt{\left(\frac{\sigma}{\mu}\right)_{\rm l}^2+
    \left(\frac{\sigma}{\mu}\right)_{\rm s}^2+
    \left(\frac{\sigma}{\mu}\right)_{\rm m}^2} \:.
\label{eq:Resolution_Combi}
\end{equation}
The additional resolution loss from the $x$-$y$-correction of the S2 signal is not included in equation \eqref{eq:Resolution_Combi} and will further increase $\left(\frac{\sigma_E}{E}\right)_{S2}$ in dual-phase and, to an expected lesser extent, in single-phase.
An accurate resolution estimate based on equation \eqref{eq:Resolution_Combi} requires the three contributing fluctuations to be normally distributed and independent. 
Neither of these conditions is fulfilled exactly. 
To evaluate the accuracy of this analytic estimator we calculate the resolution from full simulation: For each energy deposit $N_{e,0}$ is randomly determined by NEST. Each electron is then individually drifted and extracted, or not. Finally the signal generated by the $N_e$ surviving electrons is measured and the combined fluctuation determined from $10^5$ samples per energy and configuration. This ensures proper representation of the dependence between the processes and uses the exact, non-normal statistics.
Since \eqref{eq:Resolution_Combi} has an implicit non-linear dependence on $z$, the comparison, shown in Figure~\ref{fig:E_res_Comp}, treats the conservative case of events at the bottom of the TPC.
Both methods agree provided a sufficiently large number of initially liberated electrons, in-line with a reduced deviation from normal distributed fluctuations for larger $N_{e,0}$. 
For small $N_{e,0}$ equation \eqref{eq:Resolution_Combi} overestimates the resolution. 
The discrepancy between estimate and simulation is largest for high electron losses and low $N_{e,0}$, where the Poisson approximation of the binomial process fails. 
For large electron lifetimes and 100\% extraction efficiency the resolution approaches the limit set by electron liberation after the electron recoil (grey in Figure~\ref{eq:Resolution_Combi}). 
In this case the resolution deteriorates predominantly due to fluctuations in the signal measurement.

\begin{figure}[htb]
\begin{minipage}[c]{0.5\textwidth}          
    \includegraphics[width=\textwidth]{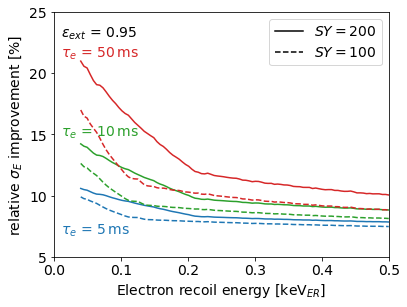}
\end{minipage}
\begin{minipage}[c]{0.5\textwidth}  
    \includegraphics[width=\textwidth]{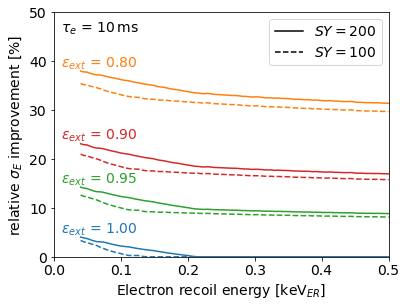}
\end{minipage}
    \captionof{figure}{Relative improvement in energy resolution  $1-\sigma_{1p}/\sigma_{2p}$ for a single-phase detector ($\sigma_{1p}$) compared to a dual-phase setup ($\sigma_{2p}$), for a fixed (varied) extraction efficiency $\epsilon_{\rm ext}$ and varied (fixed) electron lifetime $\tau_e$ in the left (right) panel. The improvement due to electron counting is limited to low energies and less pronounced for $SY = 100\,\gamma\,/\,e^-$ (dashed) than for $SY = 200\,\gamma\,/\,e^-$ (solid lines). 
    } 
    \label{fig:EnergyRes_improvement}
\end{figure}

Figure~\ref{fig:EnergyRes_improvement} shows the relative improvement in S2 energy resolution $\sigma_{E}$ for a single-phase detector with electron counting and no liquid-gas interface, compared to a dual-phase setup with otherwise identical operation conditions. 
The resolution improvement due to electron counting is limited to low energies corresponding to the $N_{e} \leq 15$ ($\leq 10$) limits for $SY = 200$ (100) photons per electron. A lower scintillation yield results in a reduced improvement, as expected. 
The magnitude of the improvement also depends on the electron lifetime (left panel). 
A purer target leads to a larger impact of measurement fluctuations and thus a higher benefit from the more accurate electron counting. 
For larger energies, the resolution improvement is solely caused by the absence of the liquid-gas interface. 
Compared to a dual-phase detector with non-ideal electron extraction efficiency, the single-phase setup yields a significantly improved energy resolution (right panel).

The resolution improvement is of particular interest for S2-only searches for sub-GeV WIMP dark matter~\cite{Aprile_2019_ChargeOnly, akerib2021enhancing}, where low-energy signals are expected. Here the S1 signal often remains below detection threshold and the energy resolution is solely driven by the charge signal. 
While for nuclear recoils fluctuations in electron-ion recombination are dominant over the accuracy of the charge signal measurement, the fluctuation in electron liberation by electron recoil (grey in Figure~\ref{fig:E_res_Comp}) is on the same order as the charge measurement uncertainty (Figure~\ref{fig:ElectronCounting_Resolution}).
In a sufficiently pure target, the reduced uncertainty from the electron counting approach combined with the absence of the liquid-gas interface can yield an energy resolution which is dominated by the irreducible fluctuations in the primary ionisation and recombination process.
For nuclear recoil searches~\cite{Aprile_2019_ChargeOnly}, this primarily improves the measurement of the ER background. For electronic recoil searches of, say sub-GeV WIMPs via the Migdal-effect~\cite{Aprile_2019_subGeV} signal and background can profit equally from the resolution improvement. 

\subsection{z-coordinate reconstruction from the charge signal}
\label{Sec:z_Reco}

The short secondary scintillation pulses per electron allow for reconstruction of the electron arrival time at the anode wire.
For an approximately point-like electron cloud created at a position $z$ below the gate, the electron arrival time distribution is dominated by the longitudinal diffusion. By measuring the width of the electron arrival time distribution $\sigma_t$, the event depth can be reconstructed using \eqref{eq:Diffusion}:
\begin{equation}
    z = \frac{\sigma_t^2  v_D^3}{2 D_L} - F_\textrm{corr}(z),
    \label{eq:z_reco}
\end{equation}
where the function $F_\textrm{corr}(z)$ corrects for the non-uniform drift- and diffusion behaviour in the electrode stack region. In this work we use the true $z$-position from MC data to define $F_\textrm{corr}$. 
Experimental calibration will rely on the precise $z$-position measurement based on the drift time between the S1 and the S2 signal. 
In the S2-only analysis $z$ is then determined using an iterative approach to solve~\eqref{eq:z_reco}.

\begin{figure}
\centering
\begin{minipage}[t]{.48\textwidth}
 \centering
  \includegraphics[width=.99\linewidth]{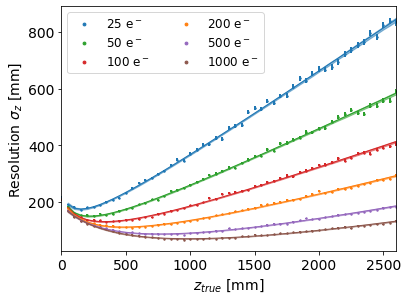}
  \captionof{figure}{Absolute $z$ resolution $\sigma_z$ in S2-only reconstruction as a function of the true event depth for different event sizes. Simulated data (points) is fitted with a 4 parameter model (line $\pm 2 \sigma$ bands), yielding $\chi^2 / {\rm dof} < 2$ for all data sets.}
  \label{fig:z_Reco_results}
\end{minipage}
\hfill
\begin{minipage}[t]{.48\textwidth}
 \centering
  \includegraphics[width=.99\linewidth]{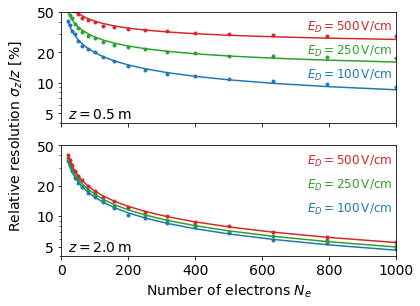}
  \captionof{figure}{Relative $z$ resolution $\sigma_z/z$ as a function of $N_{e}$ for  different drift fields and $z=\SI{0.5}{m}$ (top) and $z=\SI{2.0}{m}$ (bottom). The line indicates the $\propto 1/\sqrt{N_{e}}$ model fit, expected from the statistical fluctuation in the electron arrival time width $\sigma_t$.} 
  \label{fig:z_Reco_ED}
\end{minipage}
\end{figure}

The absolute spatial resolution $\sigma_z$ is shown for $N_e \in [25, 1000]$ in Figure~\ref{fig:z_Reco_results}. The simulated data is described by a linear function plus an additional term modelling the resolution degradation for small $z$:
\begin{equation}
    \sigma_z (z) =  p_0 z + p_1 + \frac{p_2}{\sqrt{z + p_3}}  \:.
\end{equation}
The $p_i$ are model parameters in units of mm. 
The observed resolution degradation at small $z$  arises from the $x$-$y$-position-dependence of the drift time due to the focusing effect of the gate grid, discussed in chapter~\ref{Sec:PropScint} and shown in Figure~\ref{fig:HexGrid_timing}. 
Electron clouds which are collected faster (slower) are reconstructed at smaller (larger) depth.
This $x$-$y$-dependence is not resolved by the detector and does not enter the S2-only $z$ calculation, which worsens the overall resolution. 
An alternative anode-gate alignment, as discussed in chapter~\ref{Sec:Intro}, could avoid this resolution degradation for low $z$. 
The effect is less pronounced for deeper interactions, due to the stronger longitudinal diffusion of the electron cloud. For large $z$, the resolution is determined by the statistical fluctuation in the electron arrival time and $\sigma_z$ grows linearly with $z$. 
A relative resolution $\sigma_z / z$ of 22.9\%,  16.5\% or 12.0\% is obtained for signals with 50 e$^-$ ($\approx\SI{8}{keV_\textrm{NR}}$), 100  e$^-$ ($\approx\SI{20}{keV_\textrm{NR}}$) and 200 e$^-$ ($\approx\SI{60}{keV_\textrm{NR}}$), respectively, where the conversion to energy~\cite{NEST} assumes $E_D = \SI{250}{V/cm}$ and $P_s = 1$.
Figure~\ref{fig:z_Reco_ED} shows $\sigma_z / z$ as a function of $N_{e}$ for two positions in $z$ (top, bottom) and $E_D = 100$ (blue), 250 (green) and \SI{500}{V/cm} (red). Each data set follows the $\propto 1/\sqrt{N_{e}}$ behaviour of the statistical uncertainty in $\sigma_t$. 
A lower (higher) drift field yields a more (less) precise $z$ reconstruction from the charge signal. 
This tendency is more pronounced for short drift paths.

In dual-phase TPCs the S2 width is typically used for event selection~\cite{SORENSEN201141} and the potentially-biased conversion to the space coordinate $z$ is avoided. 
Contrary to the $\sigma_t$ measured in a single-phase setup, the S2 width in a dual-phase detector is broadened by the electron scintillation time and thus the $x$-$y$-dependent height of the gas gap. This calls for an additional width calibration and correction, which becomes obsolete in single-phase.
Compared to the resolution in S2-width in XENON1T~\cite{XENON1T_analysis}, the data presented in Figure~\ref{fig:z_Reco_results} suggests a slightly lower accuracy. 
Simulation with $E_D = \SI{100}{V/cm}$, close to the experimental conditions of~\cite{XENON1T_analysis}, yields a resolution similar to the one reported from XENON1T. This comparability in precision as well as the observed $\propto 1/ \sqrt{N_{e}}$ dependence suggests that the dominant uncertainty in $z$, or S2 width, measurement stems from the fluctuation in the electron cloud's transverse drift, not from the scintillation process. Consequently, even precise electron arrival time determination does not significantly improve the $z$-resolution. 

The standard $z$-position reconstruction by drift time measurement in S1-S2 analyses is significantly more precise than the charge-signal-only inference. 
However, in an S2-only analysis, even a rather coarse determination of the event's $z$-position allows for fiducialisation, improved corrections and background suppression. 
In S1-S2 analyses, the charge-only inference provides an additional and independent measurement of $z$. 
Similarly to the S2-width based selection cut, this could be exploited for rejection of accidental coincidence events, where a lone-S1 and a lone-S2~\cite{XENON_analysis2} from independent processes happen within the maximum drift time.
A consistency check of both $z$ measurements constrains the allowed $z$ space and accordingly the coincidence time window. 
For example, a \SI{0.7}{keV} electronic recoil will likely create an S1 signal below detection threshold and thus a lone-S2 with on average 50 electrons. 
If this were accidentally paired with a lone-S1 signal of the right magnitude it could mimic a WIMP signal.
In the MC study the $z$-consistency test rejects  47\% of such accidental coincidence events while retaining 99\% signal acceptance. 
The rejection power further increases  to 60\% and 69\% for larger S2 signals with 100 and 200 electrons, respectively.

\subsection{Single site - multiple site discrimination}
\label{SS-MS_Disc}

\begin{figure}[htb]
\centering
\begin{minipage}[t]{.48\textwidth}
 \centering
  \includegraphics[width=0.99\linewidth]{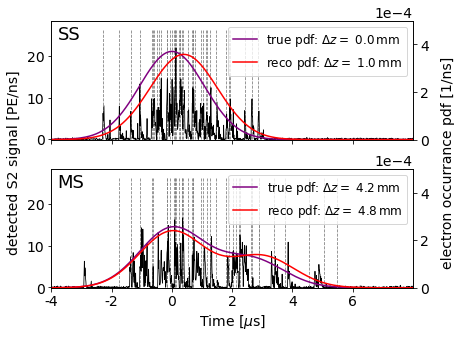}
  \captionof{figure}{Simulated single-phase waveform and reconstructed electron times (dashed grey) for single site (SS, top) and multiple site (MS, bottom) events with  40$\,e^-$ and $z=\SI{1}{m}$. The lines show the true (purple) and reconstructed (red) probability density function (pdf) for electron occurrence. }
  \vspace{6mm}
  \label{fig:SS_MS_WFexample}
\end{minipage}%
\hfill
\begin{minipage}[t]{.48\textwidth}
 \centering
  \includegraphics[width=0.99\linewidth]{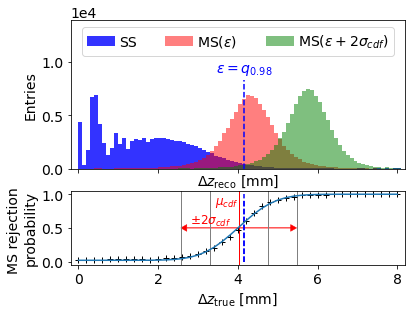}
  \captionof{figure}{(Top) Histogram of reconstructed $\Delta z$ of events with 40$\,e^-$ and $z=\SI{1}{m}$ for single site (blue) and multiple site interactions with $\Delta z_\textrm{true} = \epsilon$ (red) and $\Delta z_\textrm{true} = \epsilon + 2 \sigma_\textrm{cdf}$ (green). (Bottom) MS rejection probability vs.~true $\Delta z$, fitted by a normal cumulative distribution function (cdf).}
  \label{fig:SS_MS_Histogram}
\end{minipage}
\end{figure}

In this chapter we exploit the electron arrival time distribution to discriminate single site (SS) events, as expected for WIMP scatters, from multiple site (MS) interactions, frequently caused by neutrons. 
For MS events the $z$ difference between the two interaction locations $\Delta z$ could allow for rejection of these background events, provided $\Delta z $ can be resolved. Vice versa signal-like SS events, with true $\Delta z = 0$, must pass the discrimination with a pre-defined acceptance efficiency, e.g., 98\% used in this study. 
For discrimination, the electron arrival time distribution, shown in Figure~\ref{fig:SS_MS_WFexample}, is fitted with two Gaussian: 
\begin{equation}
    f (t) =  \frac{N_1}{\sigma_1 \sqrt{2\pi}} \exp\left(- \frac{1}{2} \left(\frac{t - \mu_1}{\sigma_1}\right)^2 \right)
    +  \frac{N_2}{\sigma_2 \sqrt{2\pi}} \exp\left(- \frac{1}{2} \left(\frac{t - \mu_2}{\sigma_2}\right)^2 \right)\: .
\end{equation}
The widths $\sigma_{1} \approx \sigma_{2}$ are fixed to the diffusion value corresponding to the S1-S2 measured drift time. The total number of electrons $N_{\rm tot} = N_1 + N_2$ is known from the signal and only the share between the presumably two interactions $N_1:N_2$ enters as free parameter, as does the mean time per signal $\mu_{1}$ and $\mu_{2}$. From the latter the parameter of interest $\Delta z = \vert \mu_{1} - \mu_{2} \vert / v_D$  is calculated for each event. To suppress false reconstruction of isolated electrons as multiple site events, we constrain $N_1, N_2 > 0.1 \, N_\textrm{tot}$ and $N_1, N_2 > 5$. 

The threshold $\epsilon$ for SS-MS discrimination is derived from simulations as the 98\% quantile $q_{\rm 0.98}$ in the true single site population, shown in blue in Figure~\ref{fig:SS_MS_Histogram} (top). Any charge signal waveform reconstructed with $\Delta z_{\rm reco} > \epsilon$ is rejected as MS event. True single site events are accepted with 98\% probability by setting the threshold $\epsilon = q_{\rm 0.98}$ . 
The rejection efficiency for multiple site events with a given $\Delta z_{\rm true}$ is shown in Figure~\ref{fig:SS_MS_Histogram} (bottom). It can be approximated by the normal cumulative distribution function (cdf), with $\mu_\textrm{cdf} \lessapprox \epsilon$ and $\sigma_\textrm{cdf} \approx \sigma_{\textrm{MS}(\Delta z = \epsilon)}$, the distribution width of MS events with true $\Delta z = \epsilon$. 
For these multiple site events with $\Delta z_{\rm true} = \epsilon$, the red population in Figure~\ref{fig:SS_MS_Histogram}  (top), the rejection probability is $> 50\%$. 
This is due to a slight positive bias in $\Delta z$ reconstruction caused by the hexagonal structure of the gate electrode.
For increasing separation in $z$, i.e., $\Delta z_{\rm true} = \epsilon + 2  \sigma_\textrm{cdf}$ (green population), the MS rejection probability rises to $ \geq 97.8\%$. 

\begin{figure}
\centering
\begin{minipage}[t]{.48\textwidth}
 \centering
  \includegraphics[width=.99\linewidth]{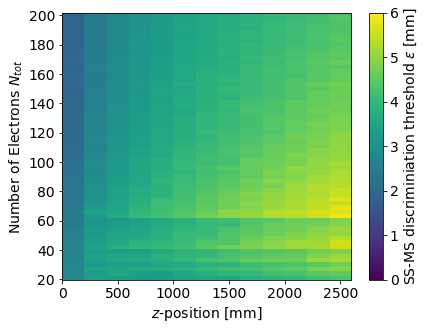}
  \captionof{figure}{Threshold $\epsilon$ for SS-MS discrimination, defined by 98\% single site acceptance, per $z$-position and number of electrons for $E_D = \SI{250}{V \per cm}$.
  The horizontal lines with increased thresholds stem from the integer steps in the required number of electrons in the smaller signal. }
  \label{fig:SS_MS_q98_thresholds}
\end{minipage}
\hfill
\begin{minipage}[t]{.48\textwidth}
 \centering
  \includegraphics[width=.99\linewidth]{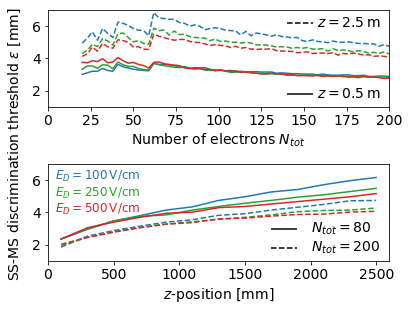}
  \captionof{figure}{SS-MS discrimination thresholds $\epsilon$ for different drift field compared along $N_{\rm tot}$ (top) and the $z$-position (bottom). Colours indicate different drift fields: $E_D = 100$ (blue), 200 (green) and \SI{500}{V/cm} (red).  }
  \label{fig:SS_MS_q98_ED}
\end{minipage}
\end{figure}

The single site acceptance thresholds $\epsilon(z, N_\textrm{tot})$ obtained from simulation are shown in Figure ~\ref{fig:SS_MS_q98_thresholds} for $z$-positions throughout the TPC, a range of $N_\textrm{tot}$ corresponding to the mean electron yield for 3~-~\SI{60}{keV} nuclear recoil events and $E_D = \SI{250}{V/cm}$. 
Throughout this parameter space a single site vs.~multiple site discrimination threshold $\epsilon \leq \SI{6.0}{mm}$ is obtained, with the lowest thresholds of \SI{1.9}{mm} for large signals at short drift distances. 
The discrimination threshold shown in Figure~\ref{fig:SS_MS_q98_thresholds} increases with larger drift lengths due to stronger longitudinal diffusion of the electron clouds and therefore increased $\sigma_t$. 
This results in a wider electron arrival time distribution for SS events and thus a larger bias in the reconstructed $\Delta z$ (blue population in Figure~\ref{fig:SS_MS_Histogram}), increasing $q_{0.98}$. 
Larger transverse diffusion increases the probability charge cloud transiting the gate through multiple openings in the hexagonal mesh.
This causes an additional separation in electron arrival time and these waveforms are more likely to be mistaken as multiple site interactions. 
A larger $N_\textrm{tot}$ decreases the significance of statistical fluctuations in the electron arrival time. Accordingly, larger signals come with improved discrimination power and lower $\epsilon$.

For a stronger drift field $E_D$ the longitudinal diffusion in the time domain $\sigma_t$ decreases. Likewise, the time separation between signals corresponding to a given spatial separation $\Delta z$ decreases, due to the higher drift velocity. 
The reduction in $\sigma_t$ has a stronger effect on the SS-MS discrimination and therefore the discrimination threshold $\epsilon$ decreases for stronger fields, as shown in Figure~\ref{fig:SS_MS_q98_ED}. 
Vice versa, lower drift fields lead to less powerful discrimination and increased thresholds. 
The difference in $\epsilon$ is small for low $z$ and increases with larger drift lengths (Figure~\ref{fig:SS_MS_q98_ED}, bottom panel). 
For $N_{\rm tot} \leq 60$ the non-continuous behaviour arising from the integer condition on the smaller signal is most pronounced for largest $z$ (top panel, dashed lines). 
Here, statistical fluctuations of a few electrons are most likely to be falsely acknowledged as second signal and reconstructed with larger $\Delta z$.

Using Geant4~\cite{Geant4_2003, Geant4_2016} and the DARWIN design and dimensions presented in~\cite{DARWIN_0vbb_2020}, we derive the distribution of distance between multiple scatter interactions as well as the deposited energies and $z$-positions for neutron MS events.
Using the interaction distance distribution, a multiple site discrimination threshold of \SI{1.9}{mm} (\SI{6.0}{mm}) can be converted to a rejection level of 95\% (87\%) of the total neutron MS background. 
The rejection rate averaged over position and energy is 93\%.
This excellent MS-SS discrimination is a clear advantage of proportional scintillation in liquid as technological choice for DARWIN. 
With this approach the DARWIN requirement of \SI{15}{mm} discrimination threshold in $z$~\cite{DARWIN_0vbb_2020} can be reached and even out-performed. 
Compared to $\epsilon = \SI{15}{mm}$ throughout the detector, the background contribution by non-resolved neutron multiple scatter events would be further reduced by a factor of 4.2, using the electron arrival time pattern produced in a single-phase TPC.

\section{Prospects of a future single-phase LXe TPC}
\label{Sec:Discussion}

Proportional scintillation in liquid xenon is a promising mechanism for charge-to-light conversion in large TPC detectors. 
Based on previous work on this subject \cite{Aprile_SinglePhase_2014} we identified geometrical and electrostatic configurations for an optimised scintillation yield with limited charge multiplication. 
The proposed experimental setup requires minimal design modifications compared to state-of-the-art dark matter detectors: an exchange of the anode, additional liquid xenon to fill above the photosensitive plane and, most critically, an increased anode voltage. 
This scheme maintains the successful detection and analysis scheme of previous and current experiments, contrary to more radical design changes~\cite{Lin_2021, SanDix_2021}.
Relying on electroluminescence in liquid xenon relaxes mechanical and electrostatic challenges related to anode wire sagging. The liquid-gas-interface and the gas gap are avoided completely. 
For ideal performance, the electrode stack geometry must be optimised in a trade-off between large $SY$, limited $f_Q$ and negligible electron emission from gate and screen. The impact of gate and screen sagging on the $SY$ uniformity can be mitigated by adjusting the inter-electrode distance and $V_A$. 
A remaining $x$-$y$-dependence of the single electron gain might require corrections, as in a dual-phase detector. 
Further S2 corrections for $x$-$y$-dependent extraction efficiency become obsolete. 
The absence of the gaseous phase furthermore allows for free orientation of the TPC towards gravity and opens up new design options like a segmented TPC with two (or multiple) amplification regions~\cite{Aprile_SinglePhase_2014, Juyal:2021njs}. 

On top of these benefits, the pulse shape scintillation signal per electron offers new analysis possibilities. 
The results presented in chapter~\ref{Sec:S2Analysis} are in first order independent of the assumed wire diameter of \SI{50}{\micro m}, provided the required scintillation yield is obtained.
Precise electron counting as quantised integer observable improves the resolution of the charge signal measurement for small signals. 
On top of this, the  energy resolution in an S2-only analysis benefits significantly from the absence of the liquid-gas-interface. 
In a single-phase setup, the S2-only energy resolution improves by more than 35\% (10\%), compared with a dual-phase setup with 80\% (95\%) electron extraction efficiency, under otherwise identical conditions.  
The measurement of the individual electron arrival times provides a charge only measurement of the $z$-position.  Albeit with limited resolution, this allows for fiducialization in the space domain. 
In a combined S1-S2 analysis a $z$ consistency check between the S2-only and the drift time based $z$ measurement yields a 50-70\% efficient discriminator against accidental coincidence events with small to medium lone-S2 signals. 
Accurate timing measurements improve the discrimination between signal-like single site and background-like multiple site events. 
Simultaneous interactions as close as \SI{6}{mm} in the drift direction can be reliably rejected throughout a \SI{40}{t} liquid xenon TPC. The false acceptance of neutron induced multiple site events as signal can be limited to well below 7\%. This significantly outperforms the DARWIN requirement on SS-MS discrimination in $z$. 

These prospects of charge signal analysis in a single-phase liquid xenon detector call for dedicated R\&D to validate the predictions presented in this simulation study. 
The key properties requiring experimental proof are the predicted sharp S2 waveforms caused by a single electron and a sufficiently high single electron gain in stable TPC operation. The scrutiny of the scintillation yield model must rely on a broader set of experimental data and is mandatory to optimise TPC design and operation conditions. Further R\&D must then focus on the scalability of this technology. 
The time line to achieve technology readiness is defined by future direct detection dark matter experiments, such as DARWIN.

\acknowledgments

We thank I.~Ostrovskiy and L.~Jones from the University of Alabama (U.S.) for the simulation of photon propagation in the proposed DARWIN single-phase detector. 
This work was supported by the European Research Council (ERC) grant No.~724320 (ULTIMATE).

\bibliographystyle{JHEP}
\bibliography{Bibliography}

\end{document}